\documentclass[preprint,12pt]{elsarticle}

\usepackage{amssymb}
\usepackage{amsmath}

\usepackage{subcaption}
\usepackage{tikz}
\usetikzlibrary{positioning,arrows.meta,calc}
\usepackage{booktabs}   
\usepackage[table]{xcolor} 
\usepackage{caption}    
\usepackage{siunitx}    
\usepackage{bm}
\usepackage{url}

\newif\ifclean
\cleantrue   

\ifclean
  \renewcommand{\textcolor}[2]{#2}
\fi

\begin{document}

\begin{frontmatter}



\title{Convolutional autoencoders for the reconstruction of three-dimensional interfacial multiphase flows}

\author[label1]{Murray Cutforth}
\affiliation[label1]{organization={Department of Mechanical Engineering, Stanford University},
            city={Stanford},
            country={USA}}

\author[label1,label2]{Shahab Mirjalili \footnote{Corresponding author}}
\affiliation[label2]{organization={Department of Engineering Mechanics, KTH Royal Institute of Technology},
            city={Stockholm},
            country={Sweden}}

\begin{abstract}
We present a systematic investigation of convolutional autoencoders for the reduced-order representation of three-dimensional interfacial multiphase flows. Focusing on the reconstruction of phase indicators, we examine how the choice of interface representation, including sharp, diffuse, and level-set formulations, impacts reconstruction accuracy across a range of interface complexities. Training and validation are performed using both synthetic datasets with controlled geometric complexity and high-fidelity simulations of multiphase homogeneous isotropic turbulence. We show that the interface representation plays a critical role in autoencoder performance. Excessively sharp interfaces lead to the loss of small-scale features, while overly diffuse interfaces degrade overall accuracy. Across all datasets and metrics considered, a moderately diffuse interface provides the best balance between preserving fine-scale structures and achieving accurate reconstructions. These findings elucidate key limitations and best practices for dimensionality reduction of multiphase flows using autoencoders. By clarifying how interface representations interact with the inductive biases of convolutional neural networks, this work lays the foundation for decoupling the training of autoencoders for accurate state compression from the training of surrogate models for temporal forecasting or input-output prediction in latent space.

\end{abstract}




\end{frontmatter}



\section{Introduction}
\label{sec:intro}

Fluid flows are considered high-dimensional dynamical systems, where real-time prediction, control, and optimization are prohibitively expensive without some form of reduced-order modeling. A low-dimensional representation of the state of a fluid system is required for a reduced-order model (ROM).  While there has been substantial progress in developing ROMs for single-phase flows \cite{rowley2017model, brunton2020machine, vinuesa2022enhancing, rozza2022advanced}, interfacial multiphase flows present a unique challenge. Multiphase flows consist of two or more immiscible fluid phases separated by a deformable interface. In this case, the state of the dynamical system strongly depends on the instantaneous location of the different phases, due to the discontinuity in density, viscosity, pressure, and other fields across the deformable phase interface. As such, it is crucial for the ROM to represent the shape of three-dimensional interfaces in multiphase flows accurately. In this work, we study the performance of convolutional autoencoders (AE) for this task. We reveal the impact of the interface representation type (diffuse, sharp, level-set function) and shape complexity on the performance of the AE. This work constitutes a robust building block for developing ROMs to predict temporal dynamics and input-output mappings in multiphase flows. Finally, considering that low-dimensional representations of complex three-dimensional shapes are highly sought after in many applications, including computer graphics, computer vision, and robotics, our findings may have implications beyond multiphase flow modeling \cite{bagautdinov2018modeling, dai2017shape, groueix2018papier}.

While substantial progress has been made in simulating (full-order modeling) interfacial multiphase flows using various approaches in the past few decades \cite{tryggvason2011direct, mirjalili2017interface}, these simulations are too expensive to allow for real-time predictions, control, or optimization. As such, ROMs for such multiphase flows are sought after. Such models are scarce in the literature \cite{wiewel2019latent, kani2019reduced, haas2020bubcnn, zhu2022review, cundy2024physics}, especially when one seeks ROMs that aim to capture the interfaces and their evolution in such flows. Here, we focus on the low-dimensional representation of the interfaces in multiphase flows via convolutional AEs. An AE is a neural network consisting of an encoder that maps the input onto a latent representation and a decoder that reconstructs the input from this representation. The encoder performs a nonlinear transformation, often reducing dimensionality, while the decoder attempts to approximate the original input as closely as possible. AEs have been successful in low-dimensional representation of single-phase flows \cite{solera2024beta, brunton2022data, kneer2021symmetry, csala2022comparing}, but their application to the reconstruction of multiphase flow states remains limited \cite{wiewel2019latent, zhu2022review}. There are multiple ways to implicitly represent interfaces in multiphase flows via field variables. These include diffuse interface representations like phase field variables, sharp interface representations like volume fractions in Volume-of-Fluid models, and level-set functions (signed-distance functions), which are inherently available in level-set-based multiphase flow models. Since one can transform these representations into one another (e.g., as a pre-processing step), we explore how the interface representation choice influences the performance of the AE. For training, we use a series of synthetic data sets with increasing shape complexity, as well as a data set consisting of snapshots from high-fidelity simulations of drops in decaying homogeneous isotropic turbulence (HIT) \cite{cundy2024physics}. This also allows us to study the effect of the complexity of the interfacial structures on the reconstruction accuracy of the AEs. 

\textcolor{green}{
While autoencoders have been increasingly adopted in fluid mechanics, existing studies have largely focused on smooth bulk fields such as pressure or velocity \cite{wiewel2019latent}, where it has been observed that small-scale features are systematically under-resolved. In interfacial multiphase flows, however, the interface geometry itself constitutes the dominant source of high-frequency content and topological complexity. The main novelty of the present work is a quantitative and systematic assessment of how the choice of interface representation influences the ability of convolutional autoencoders to represent interfacial geometry. By comparing sharp, signed-distance, and diffuse interface representations across datasets with controlled geometric complexity, we identify spectral bias as the key mechanism limiting reconstruction accuracy and demonstrate that moderately diffuse interfaces alleviate this limitation by regularizing high-wavenumber content without compromising geometric fidelity. To the best of our knowledge, this trade-off between interface representation, spectral bias, and reconstruction accuracy has not previously been established for multiphase flows.} All in all, accurately representing flow states in a low-dimensional space is a prerequisite for most ROMs. This work elucidates the best practices and limitations of applying AEs for this purpose in multiphase flows.

\begin{figure}[t]
    \centering

\begin{subfigure}[b]{0.32\textwidth}
        \centering
        \includegraphics[width=\textwidth]{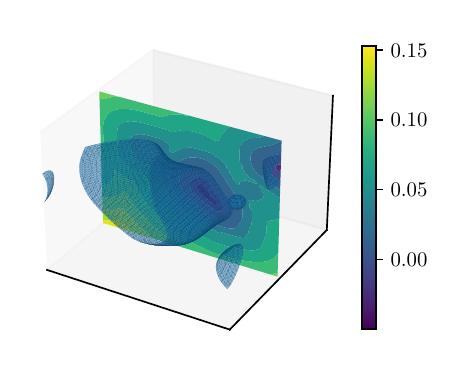}
        \caption{}
        \label{fig:1:1}
    \end{subfigure}
    \begin{subfigure}[b]{0.32\textwidth}
        \centering
        \includegraphics[width=\linewidth]{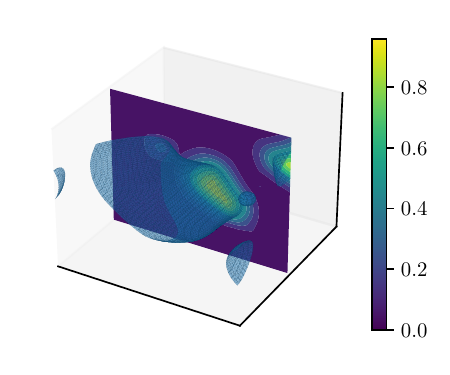}
        \caption{}
        \label{fig:1:2}
    \end{subfigure}
        \begin{subfigure}[b]{0.32\textwidth}
        \centering
        \includegraphics[width=\linewidth]{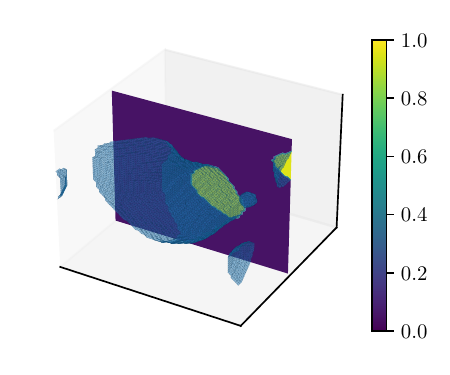}
        \caption{}
        \label{fig:1:3}
    \end{subfigure}
    
    \caption{Illustration of the three interface representations compared in this study: (a) signed-distance (level-set) functions, (b) a diffuse (tanh) interface profile, and (c) a sharp-interface indicator function, showing the interface in one snapshot of the simulation of a droplet in homogeneous isotropic turbulence.}
    \label{fig:1}
\end{figure}

\section{Methodology}

\subsection{Problem formulation}

We seek to train convolutional AEs for the reconstruction of interfacial structures in two-phase systems. Interfacial structures can be captured with various field variables. Figure \ref{fig:1} shows three different choices for full-order representation of the same interfacial structure, where panel~(a) displays the signed distance function (SDF), also known as the level-set function, typically denoted with $s$. Panel~(b) displays a diffuse interface representation using a so-called phase field variable given by
\begin{equation}
    \phi = \frac{1 + \tanh\left( \frac{s}{2\epsilon} \right)}{2},
    \label{eqn:pf_s}
\end{equation}
where $\epsilon$ is the chosen interface thickness. We denote a diffuse interface representation with thickness $\epsilon$ with ``Tanh $\epsilon$''. Panel~(c) gives a sharp interface representation, with an indicator function defined as
\begin{equation}
    H = \frac{1 - \operatorname{sgn}(s)}{2}.
\end{equation}
While these different fields represent the same interface (or phase) distributions, it is clear from Figure \ref{fig:1} that they have vastly different properties. Let us denote these fields, $(s, H, \phi)$, collectively as a three-dimensional input field $x$. The autoencoder, composed of an encoder $f_{\theta_e}^e$ and a decoder $f_{\theta_d}^d$, where $\theta_e$ and $\theta_d$ denote the learnable parameters of the encoder and decoder networks, respectively, is trained to reconstruct $x$ as accurately as possible by minimizing a loss function that quantifies the reconstruction error. Two common choices for this loss function are 
the \textit{mean squared error (MSE)}:
\[
\mathcal{L}(x, \hat{x}) = \|x - \hat{x}\|_2^2,
\]
and the \textit{mean absolute error (L1)}:
\[
\mathcal{L}(x, \hat{x}) = \|x - \hat{x}\|_1,
\]
where \(\hat{x} = f_{\theta_d}^d(f_{\theta_e}^e(x))\) is the reconstructed output.
The training process therefore amounts to solving the optimization problem
\begin{equation}
\min_{\theta_e, \theta_d}  \mathcal{L}\left(x, f_{\theta_d}^d \left( f_{\theta_e}^e(x) \right) \right),
\end{equation}
where $\mathcal{L}(\cdot,\cdot)$ denotes the chosen loss function. Note that $f_{\theta_e}^e(x) = z$ is the latent representation of the input, and the compression ratio is defined as the ratio of the dimensionality of the input (or output) field to that of the latent variable.


\subsection{Dataset}
\label{sec:dataset}

\begin{figure}[t]
    \centering

\begin{subfigure}[b]{0.24\textwidth}
        \centering
        \includegraphics[width=\textwidth]{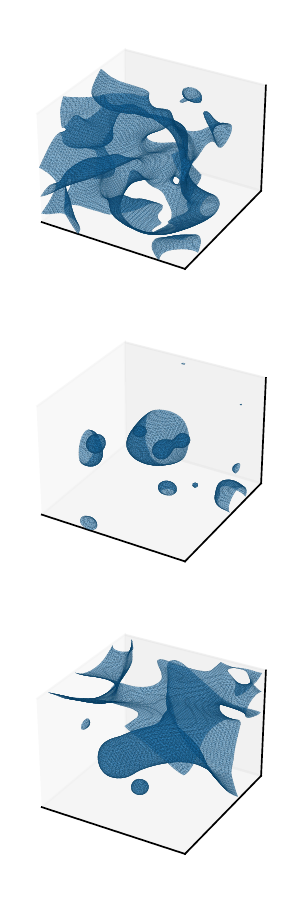}
        \caption{HIT simulation}
        \label{fig:2:1}
    \end{subfigure}
    \begin{subfigure}[b]{0.24\textwidth}
        \centering
        \includegraphics[width=\linewidth]{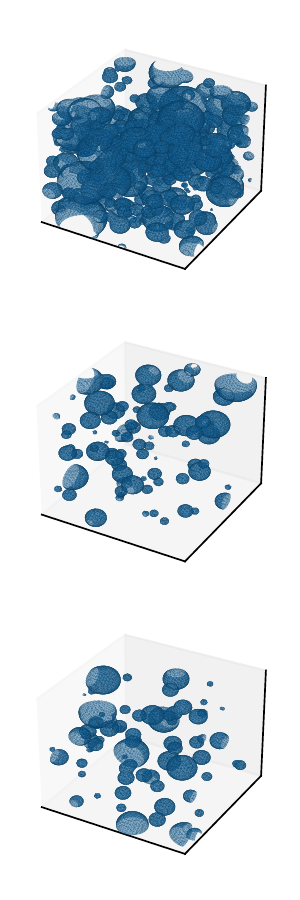}
        \caption{Synthetic ($\mu=1$)}
        \label{fig:2:2}
    \end{subfigure}
        \begin{subfigure}[b]{0.24\textwidth}
        \centering
        \includegraphics[width=\linewidth]{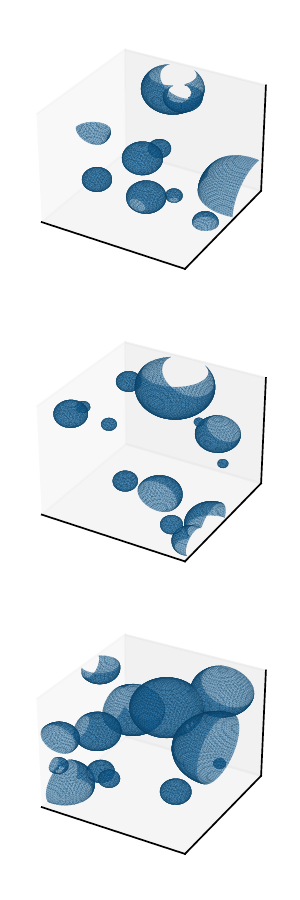}
        \caption{Synthetic ($\mu=2$)}
        \label{fig:2:3}
    \end{subfigure}
        \begin{subfigure}[b]{0.24\textwidth}
        \centering
        \includegraphics[width=\linewidth]{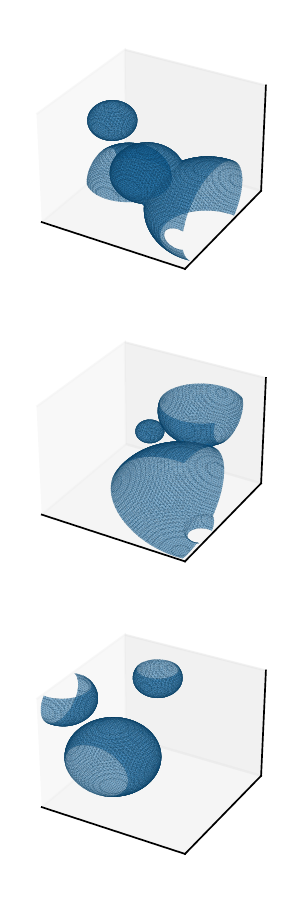}
        \caption{Synthetic ($\mu=2.5$)}
        \label{fig:2:4}
    \end{subfigure}

    \caption{Visualization of the interfacial contour for 3 samples from each of the four datasets considered in this work. In (a) the HIT simulation dataset is shown, while the remaining three panels show the synthetic dataset which is parameterized by $\mu$. $\mu=1$ in (b), $\mu=2$ in (c), and $\mu=2.5$ in (d). $\mu$ parametrizes these datasets through Equation \ref{eq:mu}.}
    \label{fig:2}
\end{figure}

Two datasets are used in this work. Each dataset is divided into a training split, a test split, and a hyper-parameter validation split in a 80/15/5 ratio.
\subsubsection{Interfacial flow simulation dataset} To incorporate a realistic flow, we use snapshots from simulations of drops in decaying homogeneous isotropic turbulence (HIT) that we previously used for training a machine learning (ML) model for drop breakup prediction~\cite{cundy2024physics}. 

\textcolor{blue}{To generate the HIT dataset in ~\cite{cundy2024physics}, direct numerical simulations (DNS) of drops in decaying homogeneous isotropic turbulence are performed using our in-house two-phase solver based on the conservative Allen--Cahn phase field method \citep{mirjalili2020conservative,mirjalili2021consistent}. The governing parameters are $\rho_1=\rho_2=5$, $\mu_1=\mu_2=0.005$, and $\sigma=0.025$ in SI units, corresponding to unity density and viscosity ratios. The resulting flows are characterized by a small Ohnesorge number and Weber numbers in the range $\mathcal{O}(0.1)$--$\mathcal{O}(100)$, which produce a wide variety of drop deformation and breakup behaviors relevant to emulsions. All simulations are conducted in a periodic domain of size $1 \times 1 \times 1$ on a uniform staggered grid of $256^{3}$ cells. The velocity field is initialized by running a single-phase forced HIT simulation for $\Delta t = 2$ using the linear forcing method of Rosales and Meneveau (2005), with the forcing coefficient $B$ sampled from $U(1, 6.5)$. This produces flows with $Re_{\lambda} \sim \mathcal{O}(10)$ while ensuring that the Kolmogorov scale is resolved. After the forcing is removed, droplets with randomized geometry and position are inserted, and the two-phase system is evolved under decaying turbulence until $t_{\mathrm{final}} = 2$. Because the density and viscosity ratios are unity, drop insertion perturbs the momentum field only through surface tension forces, enabling stable two-phase DNS. This workflow allows us to generate on the order of $10^{4}$ independent simulations. For each simulation, $5$ snapshots are saved.} 

Patches of size $64^3$ are extracted from these snapshots, by randomly choosing 64 patch locations per volume and discarding empty patches. Knowing the interface thickness ($\epsilon_\text{sim}$) used in the simulations, we compute the SDF ($s$ in Section~2, by inverting Eq.~\eqref{eqn:pf_s}), from which we construct input fields with various diffuse interface thicknesses, given by
\begin{equation}
    \phi = \frac{1 + \tanh\left(\frac{s}{2\epsilon}\right)}{2},
\end{equation}
with the limit of $\epsilon \to 0$ giving us the sharp interface representation. \textcolor{blue}{A total of 25,000 patches are extracted from this dataset.}

\subsubsection{Synthetic dataset}
To study the effect of interface complexity, we construct a series of three synthetic datasets, which are illustrated in Figure~\ref{fig:2}. We parameterize the data generation using a single variable, denoted by $\mu$, which controls the droplet size distribution and number of droplets per sample. In this way, as $\mu$ decreases, we obtain synthetic datasets with progressively increasing topological and geometrical complexity, and finer spatial scales. Each sample is constructed by taking the union of $N_D$ spherical droplets, where the droplet radii $r$ are drawn independently from
\begin{equation}
\label{eq:mu}
    r \sim \mathrm{lognormal}(\mu,\, 1/2).
\end{equation}
\textcolor{blue}{
The number of droplets $N_D$ is drawn from a uniform distribution, $N_D \sim U(1, N_{\max})$, where $N_{\max}$ is determined using a reference droplet volume based on the expected radius $\bar{r}=\exp(\mu+0.25)$. Specifically, defining a reference volume $V=\frac{4\pi}{3}\bar{r}^3$, we choose $N_{\max}$ such that $N_{\max}V=0.25$. As a result, the realized volume fraction varies stochastically between samples and is not enforced exactly, but spans a broad range up to approximately $0.25$.}
All samples are $64^3$.
\textcolor{blue}{A total of 10{,}000 volume patches are generated for each value of $\mu$ in the synthetic datasets.}

\subsection{Neural network architecture}

\begin{figure}

\centering

\begin{tikzpicture}[
    block/.style={rectangle, draw, fill=white, align=center, minimum width=10mm, minimum height=9mm, text centered, text=black, font=\sffamily},
    myarrow/.style={-Stealth, thick},
    plus/.style={circle, draw, fill=white, minimum size=6mm, inner sep=0pt, font=\LARGE\bfseries}
]

\node[block] (b1) {Weight standardized\\3$\times$3$\times$3 Conv3D};
\node[block, right=4mm of b1] (b2) {GroupNorm};
\node[block, right=4mm of b2] (b3) {SiLU};

\coordinate[right=10mm of b3] (merge);

\node[block, below=8mm of b2] (sc1) {1$\times$1$\times$1 Conv3D};

\node[plus, right=8mm of b3] (plus) {+};

\draw[myarrow] ($(b1.west)+(-7mm,0)$) -- (b1.west);

\draw[myarrow] (b1) -- (b2);
\draw[myarrow] (b2) -- (b3);
\draw[myarrow] (b3) -- (plus);

\draw[myarrow] ($(b1.west)+(-7mm,0)$) |- (sc1.west); 
\draw[myarrow] (sc1.east) -| (plus.south);

\draw[myarrow] (plus.east) -- ++(8mm,0);

\node[anchor=east] at ($(b1.west)+(-7mm,0)$) {Input};
\node[anchor=west] at ($(plus.east)+(8mm,0)$) {Output};

\draw[dashed,thick] 
    ([xshift=-4mm,yshift=4mm]b1.north west) 
    rectangle 
    ([xshift=4mm,yshift=-4mm]b3.south east);

\node[anchor=west, font=\sffamily\bfseries, fill=white, inner sep=0.5pt, outer sep=0pt]
    at ([xshift=0mm,yshift=-4mm]b3.south east) {$\times$2};

\end{tikzpicture}

    \caption{Default residual block used in the autoencoder.}
    \label{fig:3}
\end{figure}

Throughout this work, we adopt a standard convolutional architecture \cite{long2015fully}\footnote{Our architecture is available at \url{https://github.com/murraycutforth/conv-ae-3d}}; our objective is to compare interface representations rather than architecture optimization. Our network architecture follows the standard encoder-decoder structure described in Equation~1 and is based on the residual block illustrated in Figure \ref{fig:3}. 
Each layer in the encoder (decoder) consists of two such blocks in series, followed by a downsampling (upsampling) operation. 
\textcolor{blue}{Downsampling in the encoder is achieved via a 3D convolution with a stride of 2, and upsampling in the decoder is performed with a 3D transposed convolution with a stride of 2. All other convolutions use a stride of 1 with padding to preserve spatial dimensions.} 
The choice of a ResNet-type architecture~\cite{he2016deep} along with weight-standardized convolution layers is advantageous in the small batch size regime~\cite{qiao2019micro}, which is necessarily the case in 3D due to the large feature maps which must be stored. 
\textcolor{blue}{\ref{app:architecture} specifies the output shape of each block throughout the network. } Given an input of shape $1 \times H \times W \times D$, we obtain a latent representation of shape $Z \times \frac{H}{2^N} \times \frac{W}{2^N} \times \frac{D}{2^N}$, where $N$ is the number of downsampling layers in the encoder. Unless otherwise specified, we use $N=4$ and $Z=4$, resulting in a compression ratio of $1024$ and approximately $5.4\times10^6$ trainable parameters.

\subsection{Training}

All models were trained using the Adam optimization algorithm \cite{kingma2014adam} and mean absolute error loss, with a batch size of 4 and a learning rate of $10^{-5}$ (see hyper-parameter study in Section \ref{sec:hyperparams}). The number of training epochs was held constant at 100 for the synthetic datasets and 15 for the HIT simulation dataset. This was confirmed qualitatively to provide sufficient convergence. \textcolor{blue}{Some exaples of training loss curves are provided in \ref{app3}.} \textcolor{green}{Each training run utilised 4 Nvidia V100 (16GB) GPUs for 12 hours in a data-parallel approach. Note that inference cost is much lower and is performed locally, taking 10s per volume on an Apple Macbook Pro M3.}

\section{Results and discussion}

\subsection{Evaluation metrics}

Each case is evaluated using a held-out test set, according to the splits described in Section \ref{sec:dataset}. Rather than standard $L^p$ error norms, we adopt two metrics from the field of image segmentation. The evaluation of a (semantic) image segmentation is highly similar to the evaluation of an interfacial geometry. In both cases, the prediction consists of an arbitrary region represented by a binary mask. We transform all predictions to a sharp interface representation before computing metrics. \textcolor{blue}{
This is done by applying a Heaviside function to the reconstructed outputs $\hat{x}$. For signed-distance-function inputs, the sharp interface is obtained as $\hat{x}_s = H(\hat{x})$, corresponding to the zero level set. For sharp and diffuse interface representations, which take values in $[0,1]$, we use $\hat{x}_s = H(\hat{x} - 0.5)$, corresponding to a threshold at the mid-point of the phase transition. The same procedure is applied consistently to the corresponding ground-truth fields.} The following two metrics offer a concise and understandable summary of the performance of this prediction.

First, the Dice coefficient~\cite{dice1945measures} provides a value in $[0,1]$ describing the overall volumetric agreement, where a value of 1 is the best possible score:
\begin{equation}
    \mathrm{DSC}(X, Y) = \frac{2\,|X \cap Y|}{|X|+|Y|}.
\end{equation}

Second, the Hausdorff distance~\cite{huttenlocher1993comparing} is a worst-case measure, analogous to the $L^\infty$ norm. It is the maximum distance between the predicted and the true interface:
\begin{equation}
    \mathrm{HD}(\Gamma_X, \Gamma_Y) = \max \left\{ 
        \sup_{x \in \Gamma_X} \inf_{y \in \Gamma_Y} |x - y|,\;
        \sup_{y \in \Gamma_Y} \inf_{x \in \Gamma_X} |x - y|
    \right\},
\end{equation}
where $\Gamma_X$ represents an interface.
We normalize this distance relative to the length scale of the domain, so a value of 1 corresponds to a distance of 64 cell widths (for a $64^3$ grid), and a value of 0 corresponds to the best possible score.

\textcolor{orange}{
Finally, we also measure the relative volume error of the dispersed phase:
\begin{equation}
    \mathrm{RVE}_{\mathrm{abs}}(X, Y) = \frac{\big||X| - |Y|\big|}{|Y|}.
\end{equation}.
}

\subsection{Training process uncertainty}

Due to computational cost constraints (each full training run requires approximately 40 GPU-hours) we are unable to repeat each result with multiple seeds (for the random weight initialization and batch order during training). Instead, we measure the effect of the random seed once, for three different interface representations and five different random seeds, using 50\% of the HIT training set for efficiency purposes. The uncertainty in the training process is summarized in Table \ref{tab:uncertainty}. Overall, these uncertainties are smaller than many of the differences between interface representation methods measured in subsequent results.

\begin{table}[ht]
\centering
\small
\begin{tabular}{lccc ccc}
\toprule
 & \multicolumn{3}{c}{Dice} & \multicolumn{3}{c}{Hausdorff distance} \\
\cmidrule(lr){2-4} \cmidrule(lr){5-7}
Interface & Mean & Std & 95\% CI & Mean & Std & 95\% CI \\
\midrule
SDF        & 0.867 & 0.025 & (0.836, 0.899) & 0.185  & 0.127  & (0.0274, 0.342) \\
Sharp      & 0.953 & 0.004 & (0.948, 0.958) & 0.0577 & 0.0090 & (0.0465, 0.0689) \\
Tanh 1/32  & 0.957 & 0.004 & (0.951, 0.962) & 0.0666 & 0.0069 & (0.0580, 0.0752) \\
\bottomrule
\end{tabular}
\caption{Measurement of training uncertainty. Mean, standard deviation and 95\% confidence interval (CI) for Dice and Hausdorff distance, based on repeats with five random seeds. 50\% of HIT simulation dataset used. \textcolor{blue}{ ${1}/{32}$ denotes the epsilon (interface width) used with the tanh representation. }}
\label{tab:uncertainty}
\end{table}

\subsection{Hyper-parameter study}
\label{sec:hyperparams}

\begin{figure}
    \centering

\begin{subfigure}[b]{0.8\textwidth}
        \centering
        \includegraphics[width=\textwidth]{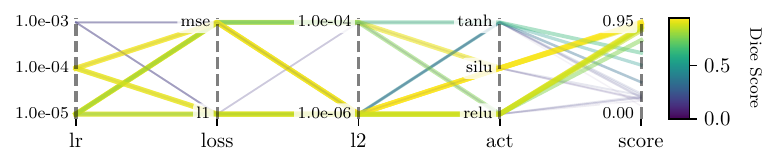}
        \caption{Sharp interface}
    \end{subfigure}


    \begin{subfigure}[b]{0.8\textwidth}
        \centering
        \includegraphics[width=\textwidth]{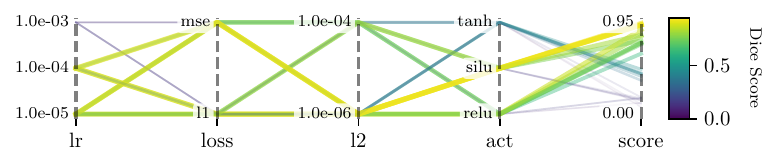}
        \caption{Tanh $\frac{1}{32}$ interface}
    \end{subfigure}


        \begin{subfigure}[b]{0.8\textwidth}
        \centering
        \includegraphics[width=\textwidth]{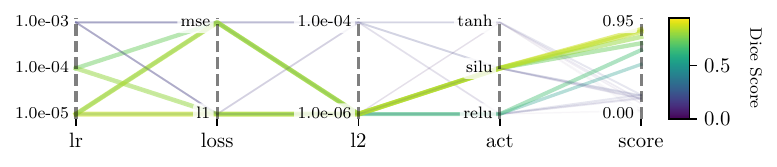}
        \caption{SDF interface}
    \end{subfigure}
    
    \caption{\textcolor{blue}{Parallel coordinate plot showing hyper-parameter grid search results for three different interface representations. Each line corresponds to a single hyper-parameter set, colored by validation set performance. Line thickness is also proportional to performance in order to highlight the best hyperparameter sets.}}
    \label{fig:hyperparams}
\end{figure}

A limited hyper-parameter grid search was run using the HIT simulation dataset. The search space consisted of learning rate, loss function, weight decay, and activation function and is summarized in Table \ref{tab:hyperparameter_search}. The dice scores on the hyper-parameter validation split for three interface types (sharp, tanh $1/32$, SDF) are summarized by a parallel coordinate plot in Figure \ref{fig:hyperparams}. 

\begin{table}[h!]
\centering
\small
\begin{tabular}{ll}
\toprule
Hyper-parameter & Grid search values \\
\midrule
Loss Function & $\bm{L_1}$, MSE \\
Learning Rate & $10^{-3}$, $10^{-4}$, $\bm{10^{-5}}$ \\
Weight Decay ($L_2$) & $10^{-4}$, $\bm{10^{-6}}$ \\
Activation Function & \textbf{SiLU}, ReLU, Tanh \\
\bottomrule
\end{tabular}
\caption{Hyper-parameter grid search space. Optimal values used in the remainder of this work highlighted in bold.}
\label{tab:hyperparameter_search}
\end{table}

\begin{figure}
    \centering
    \includegraphics[width=0.99\linewidth]{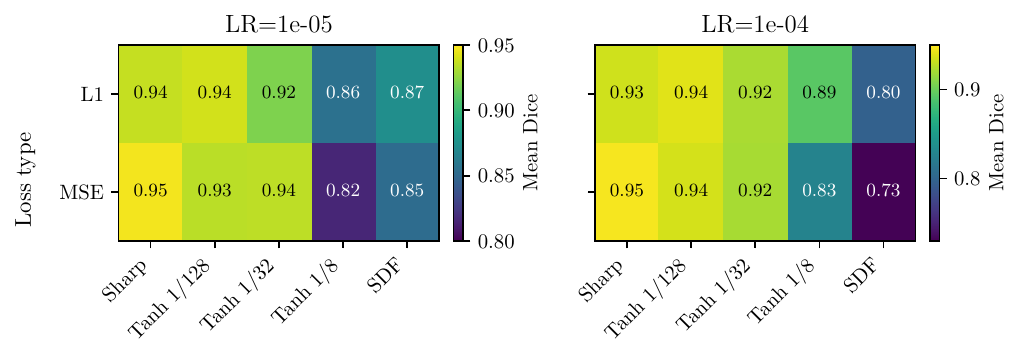}
    \caption{Hyper-parameter grid search results examining the effect of loss function for two learning rates. The $10^{-3}$ learning rate results are omitted as they did not converge. The activation function and weight decay are held at their optimal values here (\texttt{SiLU} and $10^{-6}$ respectively).}
    \label{fig:hyperparam-loss}
\end{figure}

The optimal weight decay and activation functions are unambiguously $10^{-6}$ and \texttt{SiLU}. The optimal learning rate is $10^{-5}$, but the sharp interface representation is also tolerant of a $10^{-4}$ learning rate. The choice of loss function appears to be interface-type dependent, as shown in Figure \ref{fig:hyperparam-loss}. For sharp representations, both MSE and $L_1$ are within the uncertainties estimated in the previous section. However for the more diffuse tanh interface representations and the SDF, using the $L_1$ loss offers a distinct advantage. This result may be understood in terms of MSE loss providing a smaller signal close to the interface for smoother representations.
Further details on the hyperparameter search are provided in \ref{app1}.

\subsection{Performance vs. interface representation}





    


\begin{figure}
    \centering

\begin{subfigure}[b]{0.95\textwidth}
        \centering
        \includegraphics[width=\textwidth]{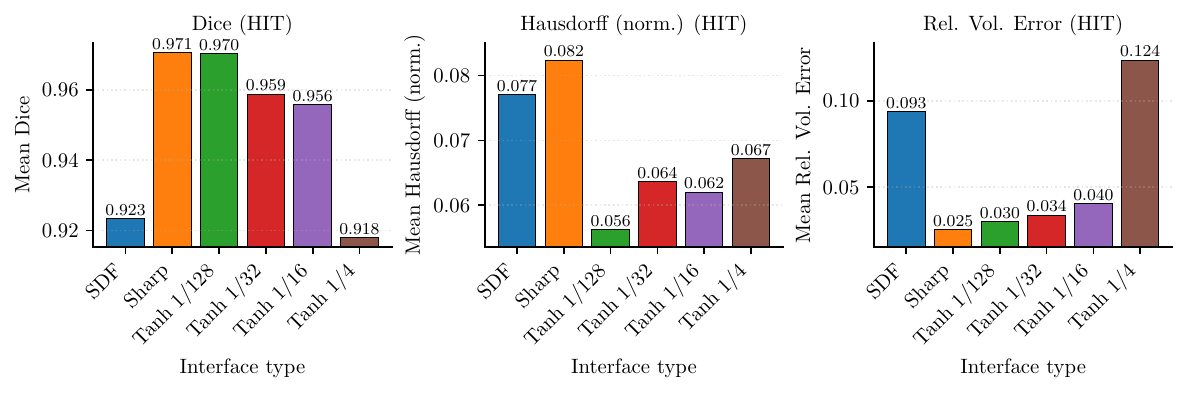}
        \caption{HIT simulation dataset}
    \end{subfigure}
    \begin{subfigure}[b]{0.95\textwidth}
        \centering
        \includegraphics[width=\textwidth]{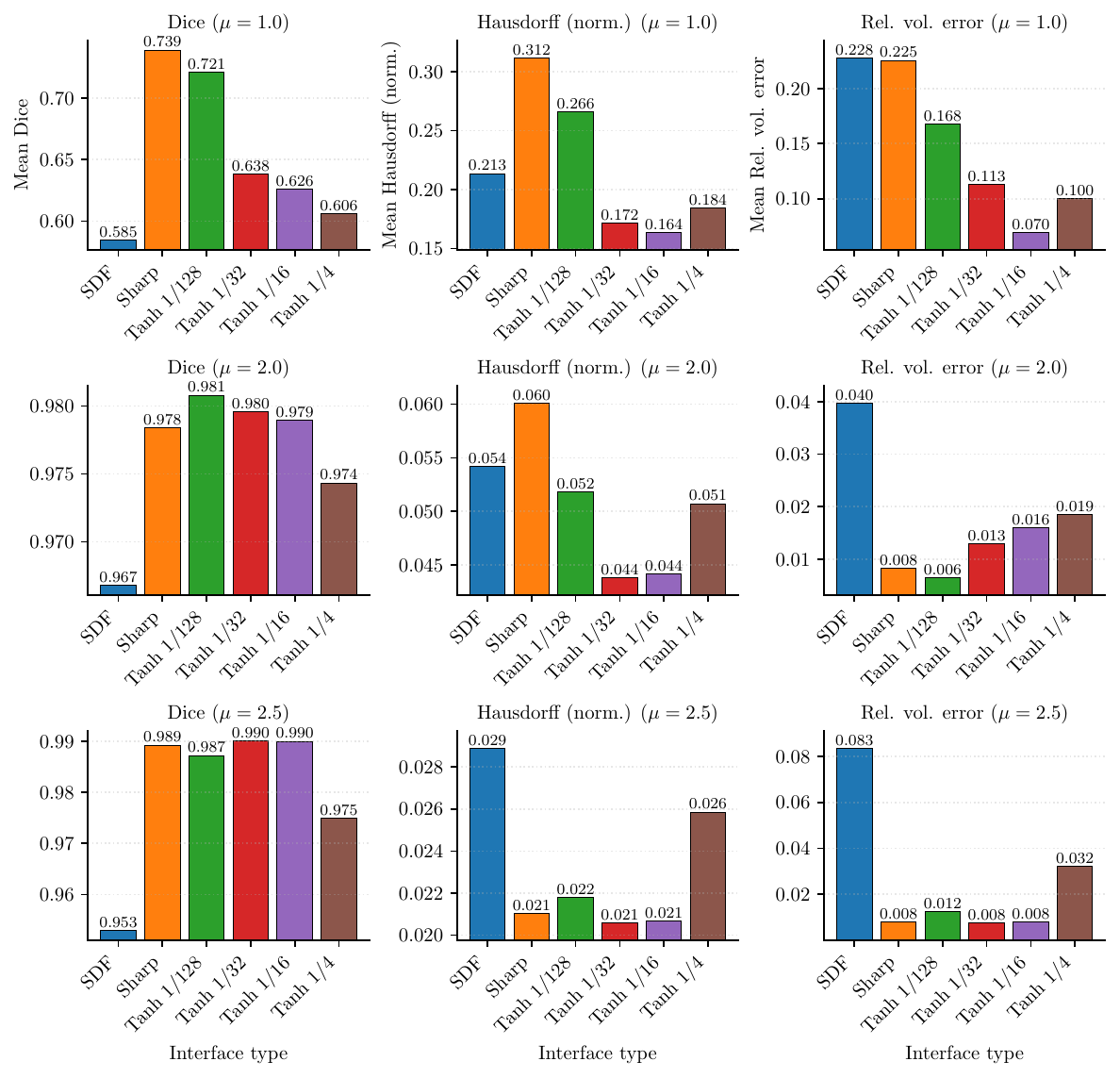}
        \caption{Synthetic datasets}
    \end{subfigure}

    \caption{\textcolor{blue}{Performance of each interface representation on (a) HIT simulation dataset and (b) synthetic datasets. Recall that for Dice, higher is better, while for Hausdorff and relative volume fraction error, lower is better. It was computationally infeasible to repeat all experiments to measure training uncertainty, but note that a subset of these runs were repeated as shown in Table \ref{tab:uncertainty}, and these suggest that these differences are significantly greater than the training uncertainty.} }
    \label{fig:comparison_hit_abs}
\end{figure}

\begin{figure}
    \centering

    
    \begin{subfigure}[b]{0.99\textwidth}
        \centering
        \includegraphics[width=\textwidth]{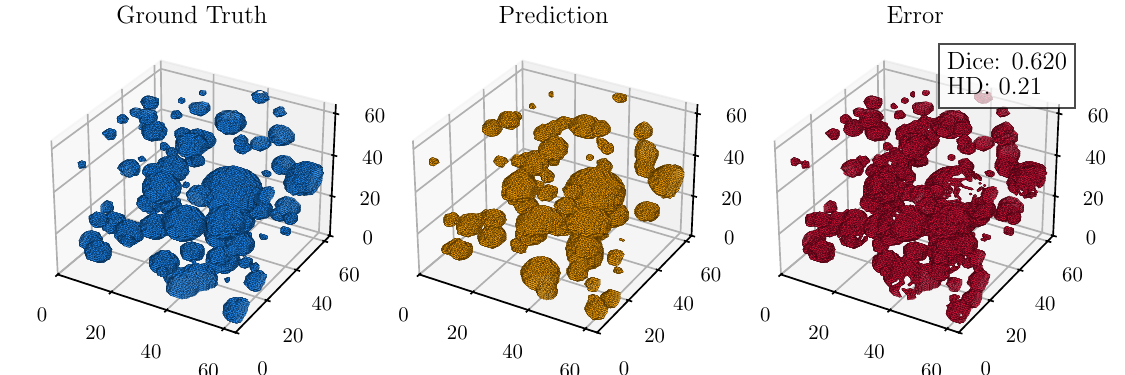}
        \caption{Tanh $\frac{1}{32}$}
    \end{subfigure}

    \begin{subfigure}[b]{0.99\textwidth}
        \centering
        \includegraphics[width=\textwidth]{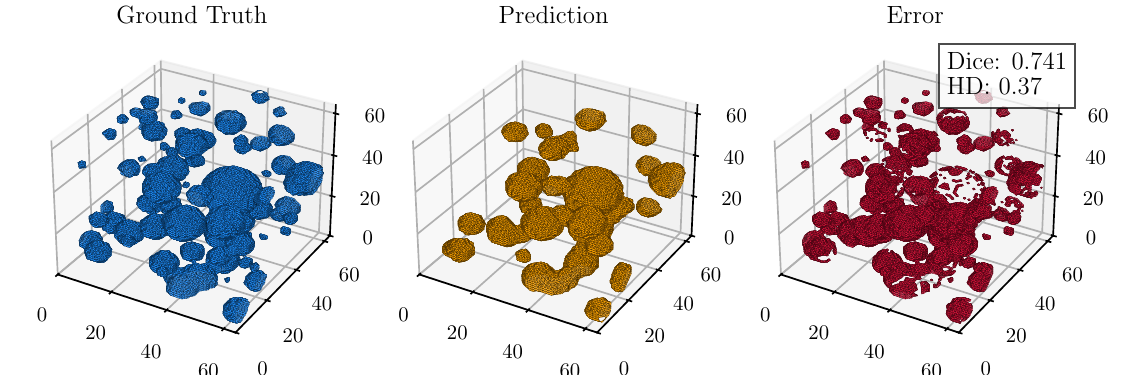}
        \caption{Tanh $\frac{1}{128}$}
    \end{subfigure}
    
    \caption{Visualization of a single test set case from the synthetic ($\mu=1$) dataset with (a) smoother interface and (b) sharper interface.}
    \label{fig:examples}
\end{figure}

Figure \ref{fig:comparison_hit_abs} shows the reconstruction performance for each \textcolor{blue}{test} dataset using various interface representations. The Dice coefficient is generally lowest for the SDF inputs and steadily improves as the interface becomes sharper, plateauing at a dataset-dependent interface thickness $\epsilon$.

While the Dice metric favors sharper interfaces, a different trend emerges when evaluating the Hausdorff distance. For more complex fields containing small drops, the sharpest representations no longer yield the best performance. Instead, a moderately diffuse interface representation tends to give the lowest Hausdorff error. A representative example shown in Figure \ref{fig:examples} illustrates this effect: sharper interfaces enable more accurate reconstruction of large-scale features, but at the expense of small-scale ones. We attribute this trade-off to the spectral bias of deep neural networks \citep{rahaman2019spectral}, which makes high-frequency features (such as sharp interfaces and fine structures) more difficult to learn during training. \textcolor{orange}{This interpretation is quantitatively supported by a spectral analysis of the reconstructed fields and reconstruction errors performed in \ref{app:psd}, revealing a systematic attenuation of high-wavenumber content in the autoencoder outputs, particularly for sharp interface representations.}

\textcolor{orange}{Volume conservation errors for the dispersed phase are also shown in Figure \ref{fig:comparison_hit_abs}. For most datasets, the SDF and the smoothest diffuse interface representation exhibit the largest volume conservation errors. The magnitude of the conservation errors are dataset dependent, typically increasing with dataset complexity. Nonetheless, consistent with the other error metrics, a moderately diffuse interface representation yields comparatively small errors across all datasets.}

\begin{figure}
    \centering
    \includegraphics[width=\linewidth]{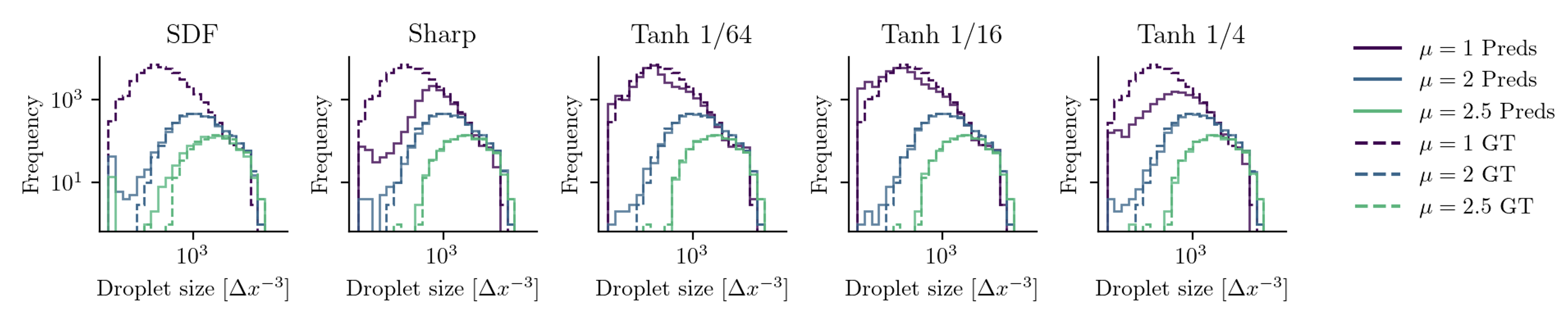}
    \caption{Comparison of predicted vs. true droplet size distributions for different interface representations. On the most complex dataset ($\mu=1$), only the tanh representation with intermediate interface thicknesses provides predictions which preserve the droplet size distribution.}
    \label{fig:droplets}
\end{figure}

The droplet size distribution is an important quantity of interest in many interfacial multiphase flows. \textcolor{blue}{We compute droplet size distributions by first converting the reconstructed fields to a sharp interface representation and then applying connected-component labeling to identify individual droplets. The volume of each droplet is computed by integrating the indicator function over each connected component.} In Figure \ref{fig:droplets}, we compare the predicted distribution of droplet size in the reconstructed samples on the synthetic dataset.  In particular, when we focus on the most complex dataset ($\mu=1$), we observe that the best match to the ground truth was given by a tanh profile with an intermediate interface thickness. This observation further supports the conclusion that moderately diffuse interfaces offer the best balance between resolving large and small features.

The underperformance of SDFs was unexpected and highlights an important distinction between interface representations. We hypothesize that this is because reconstruction errors in the SDF representation are distributed throughout the domain, rather than being localized near the interface, and thus do not penalize the model as effectively as errors in sharp or diffuse representations. Overall, our findings suggest a trade-off between sharp and diffuse representations: while sharper interfaces improve the reconstruction of large-scale features, moderately diffuse representations better capture small-scale structures in complex fields. Despite this trade-off, both sharp and diffuse representations consistently outperform SDFs. \textcolor{blue}{Finally, to assess whether nonlinear representations are essential for this task, we evaluate a linear autoencoder baseline with the same compression ratio (\ref{app:pca}). Across all datasets and interface representations, the linear model exhibits substantially larger reconstruction errors than the nonlinear convolutional autoencoder, confirming that nonlinear feature extraction is required to accurately represent complex interfacial geometries.}



\subsection{Generalization error}

\begin{figure}
    \centering
    \includegraphics[width=0.99\linewidth]{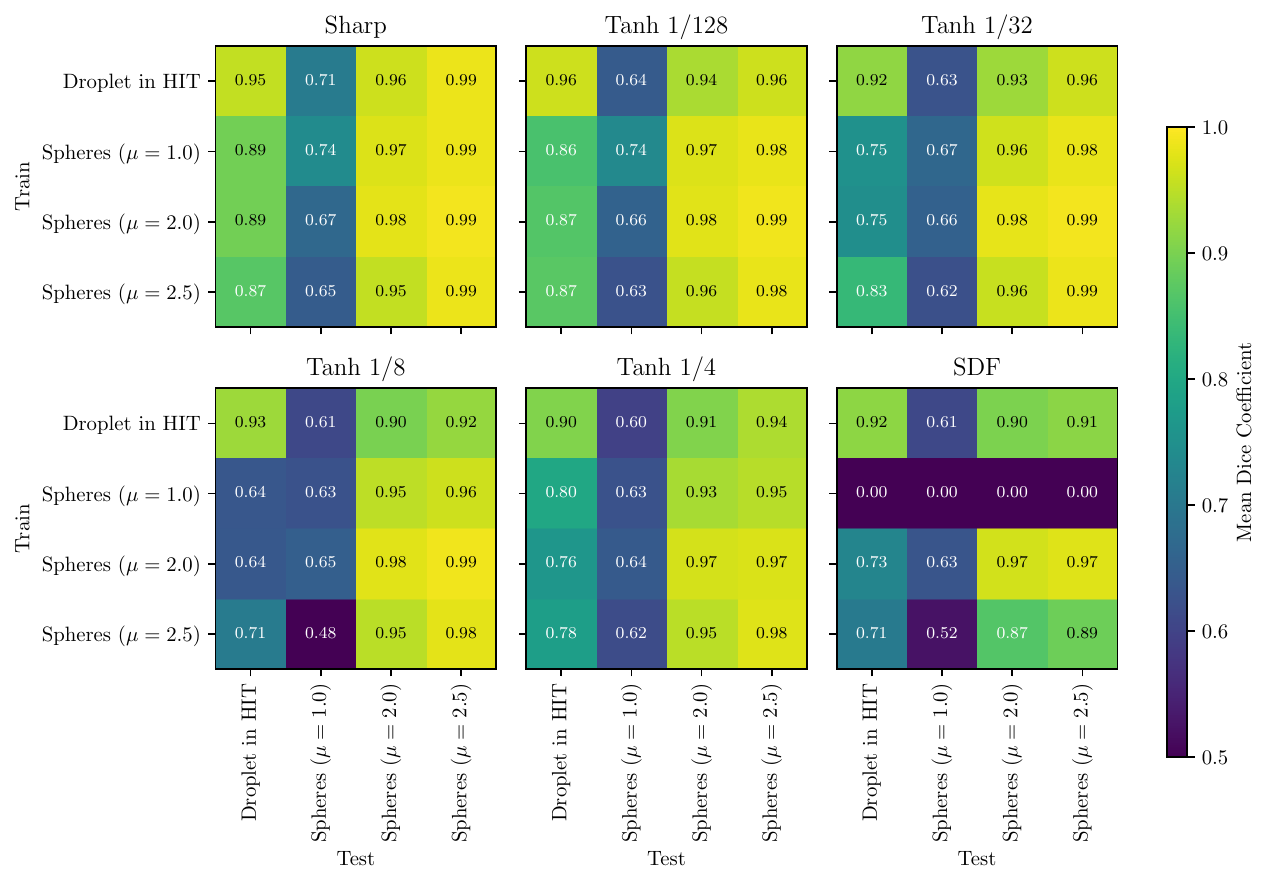}
    \caption{Comparison of generalization error for different interface representations when training and evaluating on different combinations of datasets. \textcolor{blue}{Note that zero dice coefficient indicates that training did not converge.}}
    \label{fig:ood}
\end{figure}

Finally, we investigate the generalization ability of each interface representation by evaluating the mean Dice coefficient for each combination of train/test dataset. \textcolor{blue}{For computational reasons, only 10\% of each test was used, so the absolute values of the dice coefficient are not directly comparable to Figure \ref{fig:comparison_hit_abs}.} Figure \ref{fig:ood} shows that sharper interface representations transfer better to out-of-domain test datasets. Furthermore, sharper interface models trained only on synthetic data transfer surprisingly effectively to the HIT dataset, suggesting that synthetic data augmentation is an effective tool in this domain. \textcolor{blue}{Moreover, the SDF is not only the least accurate, but also suffers from training instabilities for the most complex synthetic dataset ($\mu=1$). Specifically, the row of zeros in bottom right subfigure indicates a lack of training convergence.}

\section{Conclusions}

We have evaluated the performance of a fully convolutional three-dimensional autoencoder for reconstructing interfacial geometries using common implicit interface representations employed in multiphase simulations. Comparisons across simulated and synthetic datasets of varying complexity show that the signed-distance function consistently yields inferior reconstruction performance. In contrast, sharp and diffuse interface representations exhibit a clear trade-off: sharper interfaces improve the reconstruction of large-scale features, but excessively sharp representations tend to miss small-scale structures, while overly diffuse interfaces degrade overall accuracy.

Across all datasets and metrics considered, a moderately diffuse interface representation with an interface width of one to four grid spacings provides the best balance, preserving small-scale features while maintaining high reconstruction fidelity. We further demonstrate that this trade-off can be explained by the spectral bias of convolutional neural networks. \textcolor{orange}{Spectral analysis of the reconstructed fields confirms that high-wavenumber content associated with sharp interfaces is systematically under-represented, leading to increased errors at small scales. Moderately diffuse interfaces mitigate this effect by regularizing high-frequency content, resulting in more accurate reconstructions across spatial scales.}

\textcolor{green}{Due to the high computational cost of training three-dimensional autoencoders, this study considered a single compression ratio and a single architecture type, and training uncertainty was only partially quantified.} Future work will focus on incorporating conservation constraints into the network output, exploring variational autoencoders, and including additional state variables relevant to multiphase flows. The insights gained from this analysis lay the groundwork for developing reduced-order models capable of predicting temporal dynamics in complex multiphase systems.

\section*{Acknowledgements}

We acknowledge financial support from the US Department of Energy’s National Nuclear Security Administration via the Stanford PSAAP-III Center (DE-NA0003968). S. M. also acknowledges start-up funding from KTH Royal Institute of Technology and the Swedish e-science Research Center.

\bibliographystyle{elsarticle-num} 
\bibliography{library.bib}

\appendix

\section{Model architecture}
\label{app:architecture}

See table \ref{tab:arch}.

\begin{table}[ht]
\centering
\caption{Detailed network architecture. An input volume of shape $1 \times H \times W \times D$ is processed. ResBlocks contain SiLU activations, and all convolutions use padding to preserve dimensions unless strided. The final decoder layer is linear. $Z$ denotes the number of latent channels (default is 4).}
\label{tab:arch}
\begin{tabular}{@{}llc@{}}
\toprule
\textbf{Component} & \textbf{Layer} & \textbf{Output Shape} \\ \midrule
\multicolumn{3}{c}{\textbf{Encoder}} \\ \midrule
Input & - & $1 \times H \times W \times D$ \\
Initial Conv & Conv3d(1, 32) & $32 \times H \times W \times D$ \\ \midrule
Stage 1 & 2 $\times$ ResBlock(32) & $32 \times H \times W \times D$ \\
 & Downsample Conv3d(32, 64) & $64 \times \frac{H}{2} \times \frac{W}{2} \times \frac{D}{2}$ \\ \midrule
Stage 2 & 2 $\times$ ResBlock(64) & $64 \times \frac{H}{2} \times \frac{W}{2} \times \frac{D}{2}$ \\
 & Downsample Conv3d(64, 128) & $128 \times \frac{H}{4} \times \frac{W}{4} \times \frac{D}{4}$ \\ \midrule
Stage 3 & 2 $\times$ ResBlock(128) & $128 \times \frac{H}{4} \times \frac{W}{4} \times \frac{D}{4}$ \\
 & Downsample Conv3d(128, 256) & $256 \times \frac{H}{8} \times \frac{W}{8} \times \frac{D}{8}$ \\ \midrule
Stage 4 & 2 $\times$ ResBlock(256) & $256 \times \frac{H}{8} \times \frac{W}{8} \times \frac{D}{8}$ \\
 & Downsample Conv3d(256, 256) & $256 \times \frac{H}{16} \times \frac{W}{16} \times \frac{D}{16}$ \\ \midrule
Bottleneck & 4 $\times$ ResBlock(256) & $256 \times \frac{H}{16} \times \frac{W}{16} \times \frac{D}{16}$ \\
& Conv3d(256, 256) & $256 \times \frac{H}{16} \times \frac{W}{16} \times \frac{D}{16}$ \\
Final Encoder Conv & Conv3d(256, Z) & $Z \times \frac{H}{16} \times \frac{W}{16} \times \frac{D}{16}$ \\ \midrule
\multicolumn{3}{c}{\textbf{Decoder}} \\ \midrule
Initial Decoder Conv & Conv3d(Z, 256) & $256 \times \frac{H}{16} \times \frac{W}{16} \times \frac{D}{16}$ \\ \midrule
Stage 1 & 2 $\times$ ResBlock(256) & $256 \times \frac{H}{16} \times \frac{W}{16} \times \frac{D}{16}$ \\
 & Upsample ConvTranspose3d(256, 256) & $256 \times \frac{H}{8} \times \frac{W}{8} \times \frac{D}{8}$ \\ \midrule
Stage 2 & 2 $\times$ ResBlock(256) & $256 \times \frac{H}{8} \times \frac{W}{8} \times \frac{D}{8}$ \\
 & Upsample ConvTranspose3d(256, 128) & $128 \times \frac{H}{4} \times \frac{W}{4} \times \frac{D}{4}$ \\ \midrule
Stage 3 & 2 $\times$ ResBlock(128) & $128 \times \frac{H}{4} \times \frac{W}{4} \times \frac{D}{4}$ \\
 & Upsample ConvTranspose3d(128, 64) & $64 \times \frac{H}{2} \times \frac{W}{2} \times \frac{D}{2}$ \\ \midrule
Stage 4 & 2 $\times$ ResBlock(64) & $64 \times \frac{H}{2} \times \frac{W}{2} \times \frac{D}{2}$ \\
 & Upsample ConvTranspose3d(64, 32) & $32 \times H \times W \times D$ \\ \midrule
Final Blocks & 4 $\times$ ResBlock(32) & $32 \times H \times W \times D$ \\
 & Conv3d(32, 32) & $32 \times H \times W \times D$ \\
Final Decoder Conv & Conv3d(32, 1) & $1 \times H \times W \times D$ \\ \bottomrule
\end{tabular}
\end{table}

\section{Training loss curves}
\label{app3}

\begin{figure}
    \centering
    \includegraphics[width=0.99\linewidth]{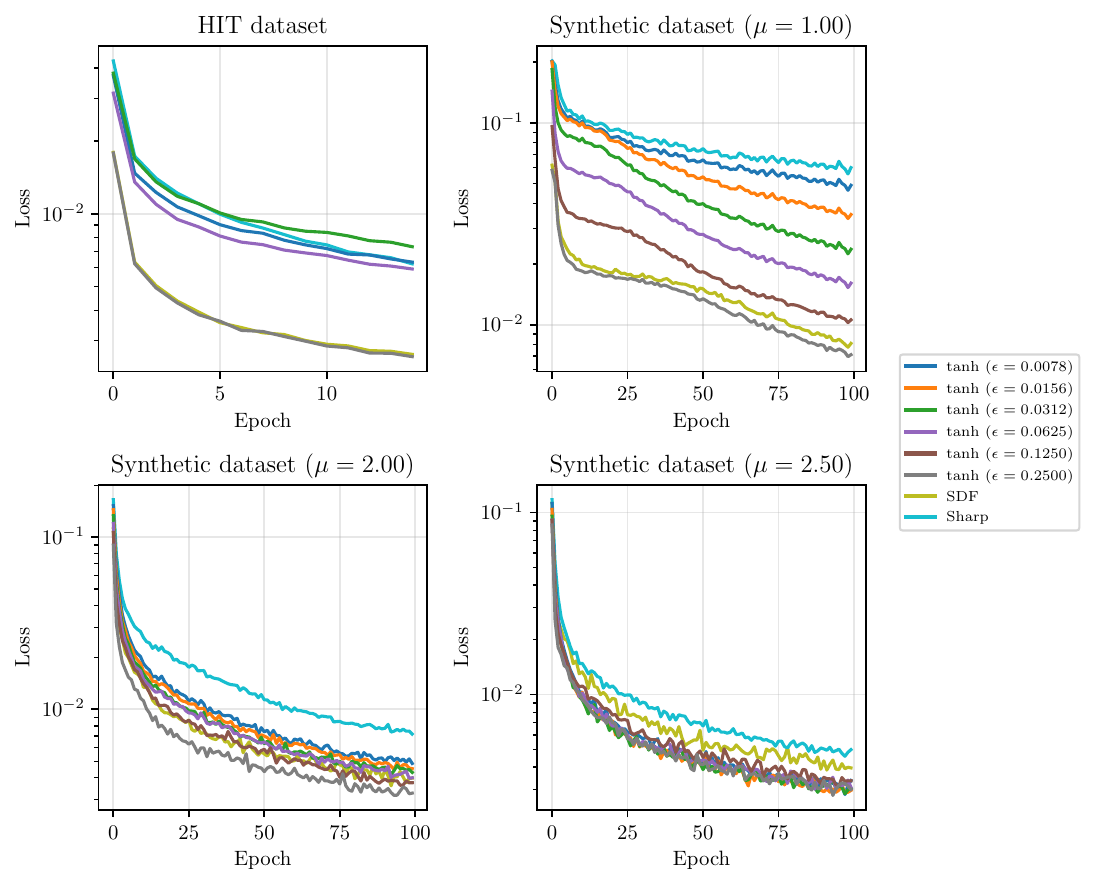}
    \caption{Training loss vs. epoch for all models compared in Figure \ref{fig:comparison_hit_abs}}
    \label{fig:loss_v_epoch}
\end{figure}

\textcolor{green}{
The training losses corresponding to the runs presented in Figure \ref{fig:comparison_hit_abs} are shown here in Figure \ref{fig:loss_v_epoch}. 
While the models have not reached complete convergence, the losses have largely plateaued, enabling meaningful comparison of their relative performance. Since the optimal learning rate was checked independently for each interface type through hyperparameter search, differences in convergence speed reflect the inherent learnability of each representation.
}

\section{Best performing hyper-parameter sets}
\label{app1}

\begin{figure}[h!]
    \centering

\begin{subfigure}[b]{0.4\textwidth}
        \centering
        \includegraphics[width=\textwidth]{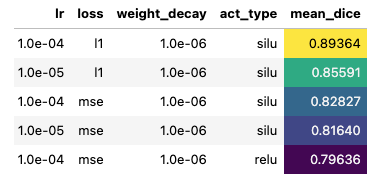}
        \caption{Tanh $\frac{1}{8}$}    
    \end{subfigure}
    \begin{subfigure}[b]{0.4\textwidth}
        \centering
        \includegraphics[width=\textwidth]{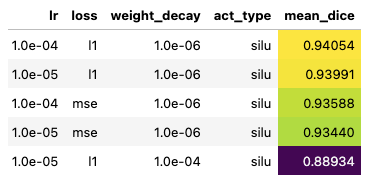}
        \caption{Tanh $\frac{1}{128}$}
    \end{subfigure}

    \begin{subfigure}[b]{0.4\textwidth}
        \centering
        \includegraphics[width=\textwidth]{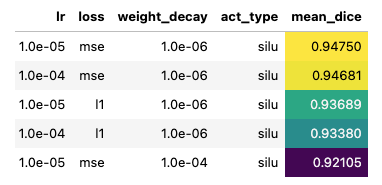}
        \caption{Sharp}
    \end{subfigure}
    \begin{subfigure}[b]{0.4\textwidth}
        \centering
        \includegraphics[width=\textwidth]{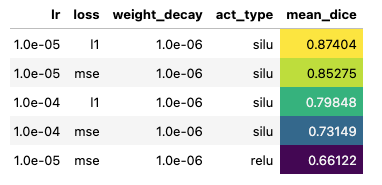}
            \caption{SDF}    
    \end{subfigure}

    \caption{Top 5 best performing hyper-parameter sets for four interface types.}
    \label{fig:top5params}
\end{figure}

We provide further detail on the optimal hyper-parameters for each interface type here. Figure \ref{fig:top5params} lists the top 5 hyper-parameter sets examined in the grid search experiment for four of the interface types.

\section{Spectral analysis of reconstruction errors}
\label{app:psd}
\begin{figure}[t]
    \centering
    \begin{subfigure}[b]{0.49\textwidth}
        \centering
        \includegraphics[width=\linewidth]{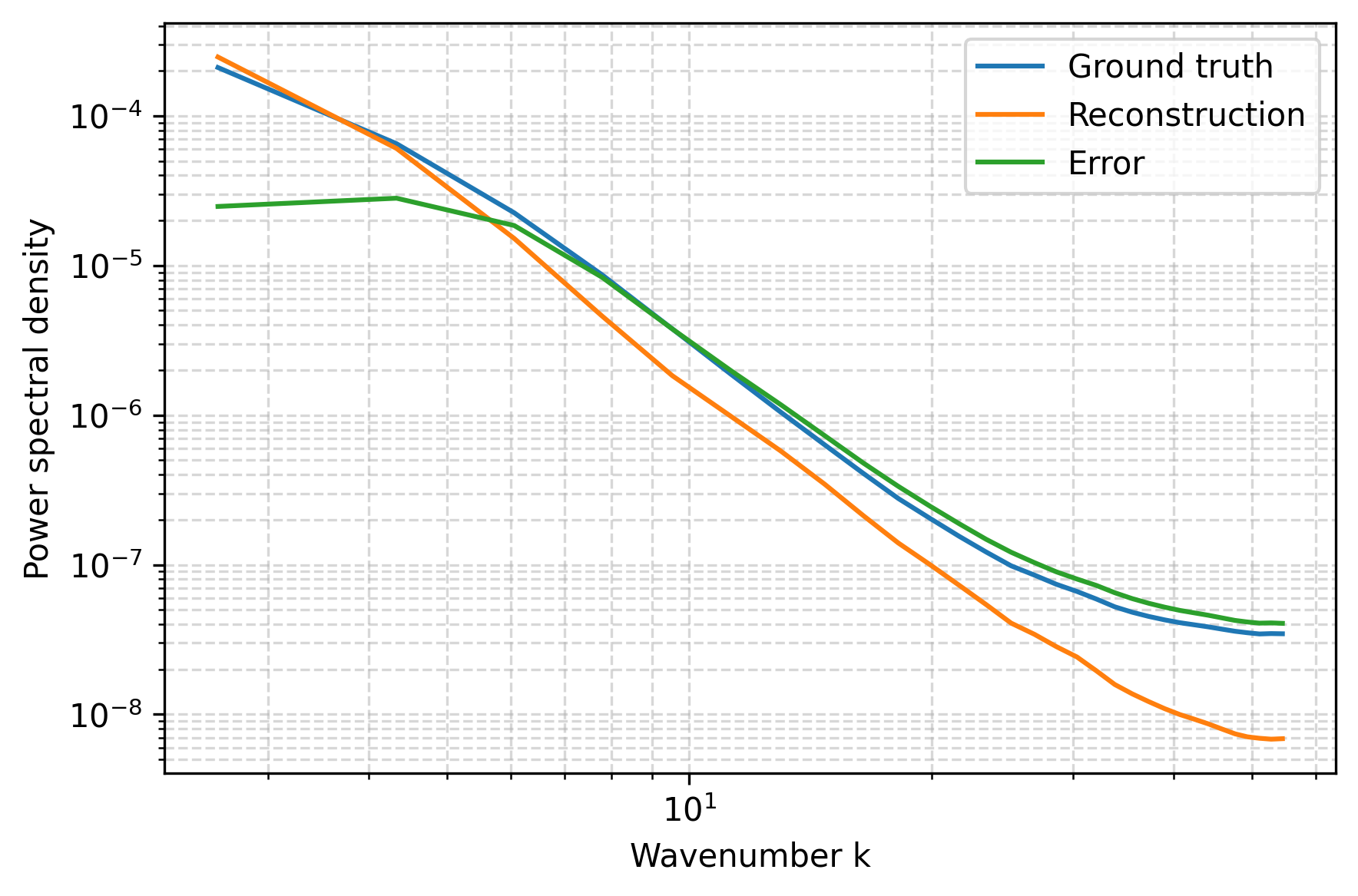}
        \caption{Sharp}
        \label{fig:psd:2}
    \end{subfigure}
        \begin{subfigure}[b]{0.49\textwidth}
        \centering
        \includegraphics[width=\linewidth]{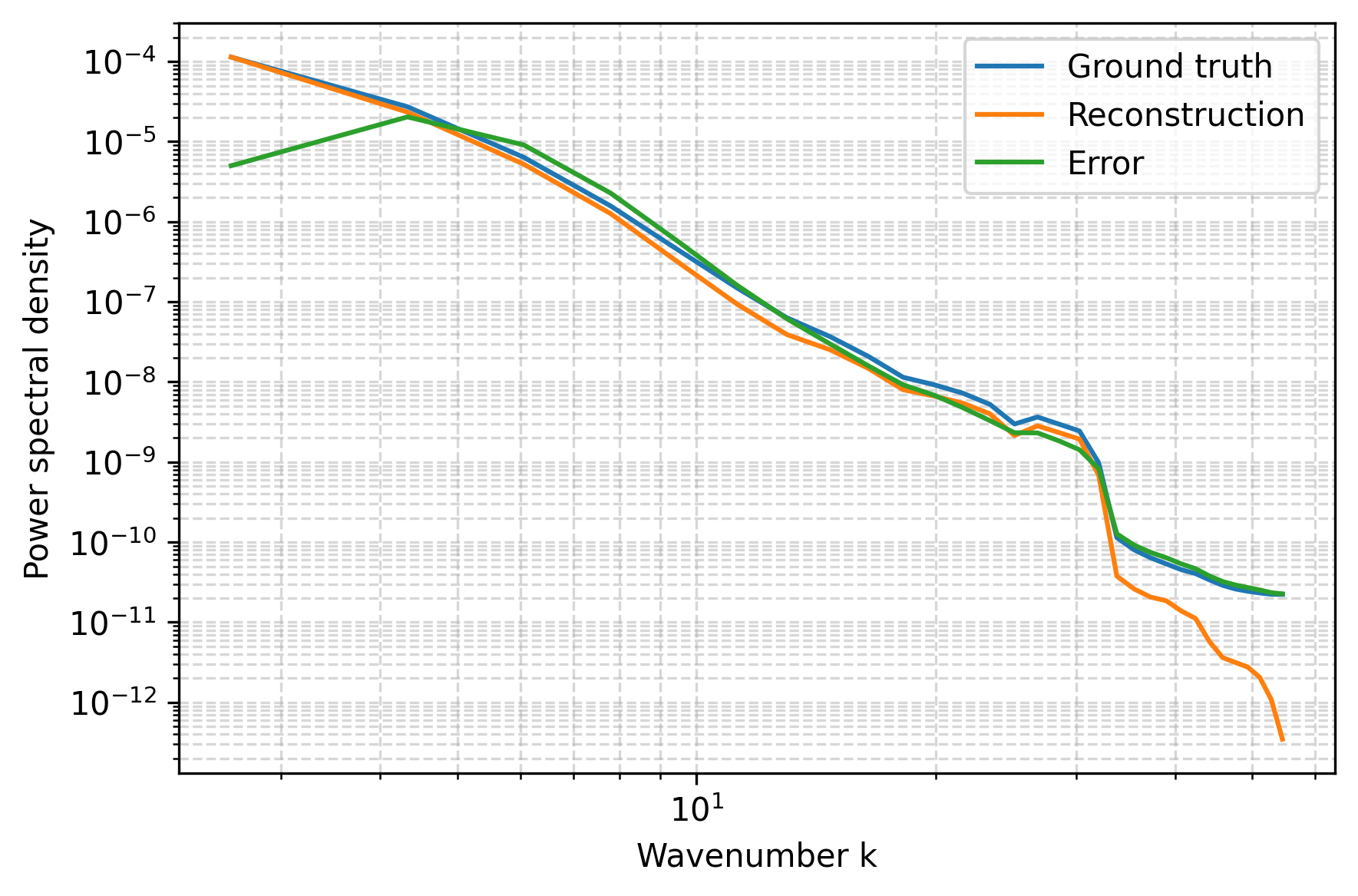}
        \caption{Tanh $1/32$}
        \label{fig:psd:4}
    \end{subfigure}
    \caption{Radially averaged three-dimensional power spectral densities for the $\mu=1$ synthetic dataset. Results are shown for (a) sharp interface and (c) diffuse interface with $\epsilon=1/32$. In each case, the ground truth, reconstruction, and reconstruction error spectra are shown. The reconstructions systematically under-represent high-wavenumber content, with the severity of attenuation depending on the interface representation.}
    \label{fig:psd_mu1}
\end{figure}

\textcolor{orange}{To quantify how reconstruction accuracy varies across spatial scales, we compute radially averaged three-dimensional power spectral densities (PSDs) of the reconstructed interface fields and corresponding ground-truth samples. For each sample, the field is first mean-subtracted and transformed using a three-dimensional Fourier transform. The power spectrum is then obtained from the squared magnitude of the Fourier coefficients and radially averaged over spherical shells in wavenumber space. All spectra are computed consistently for the ground truth, the autoencoder reconstructions, and the reconstruction error (defined as the difference between reconstruction and ground truth), and are subsequently averaged over the test set. The analysis is performed for the most challenging synthetic dataset ($\mu = 1$), where fine-scale interfacial structures are most pronounced. We focus on two representative interface formulations: a sharp interface representation and a diffuse interface representations based on a hyperbolic tangent profile with interface thicknesses $\epsilon=1/32$. Figure~\ref{fig:psd_mu1} shows the PSDs for the ground truth, reconstructions, and reconstruction errors for the two interface representations. Across both cases, the reconstructions closely match the ground-truth spectra at low wavenumbers. However, a systematic loss of energy is observed at high wavenumbers. This provides direct evidence that fine-scale features are preferentially lost in the reconstruction. Even in the ground truths (inputs to the autoencoder), the spectral content of the two interface representations are significantly different. The sharp interface representation exhibits a slow decay of energy content across wavenumbers. As a result, there is significant mismatch in the energy content of the reconstructed sharp interface fields (large errors) for moderate to high wavenumbers. On the other hand, the diffuse interface representations show strong decay of energy content across the wavenumbers. As such, the energy content of the error is much smaller than the sharp-interface representation for moderate to high wavenumbers (small scales). This analysis demonstrates the spectral bias of the autoencoder neural network, where the trained model struggles to reconstruct the small features and sharp interfaces that are dominant in the sharp interface representation.}

\section{Linear autoencoder baseline}
\label{app:pca}

\textcolor{blue}{To assess whether nonlinear representations are necessary for accurate reconstruction of interfacial multiphase flows, we consider a linear autoencoder baseline. A linear autoencoder consists of a single linear encoder and decoder without activation functions. When trained to optimality using a squared reconstruction loss, this model is mathematically equivalent to principal component analysis (PCA) \cite{baldi1989neural}. Including this baseline allows us to evaluate whether the improved performance of convolutional autoencoders arises from their nonlinear feature extraction capabilities, rather than from architectural or optimization choices alone. The linear autoencoder is trained using the same datasets, train–validation splits, and compression ratio as the nonlinear convolutional autoencoder presented in the main text. In particular, the latent dimensionality is fixed to 256 for all experiments. The model is trained using the Adam optimizer to minimize the mean squared reconstruction error. While gradient-based optimization does not guarantee convergence to the global optimum of the linear model, the resulting performance provides a practical and relevant baseline for comparison under identical training conditions. Across all datasets and interface representations considered, the linear autoencoder exhibits substantially larger reconstruction errors than the nonlinear convolutional autoencoder. This performance gap is observed consistently for sharp, diffuse, and signed-distance interface representations. These results demonstrate that linear dimensionality reduction techniques such as PCA are insufficient for accurately representing complex three-dimensional interfacial geometries in multiphase flows, and confirm that nonlinear autoencoders provide a necessary and meaningful advantage for reduced-order representation in this setting.}


\begin{figure}
    \centering
    \includegraphics[width=0.99\linewidth]{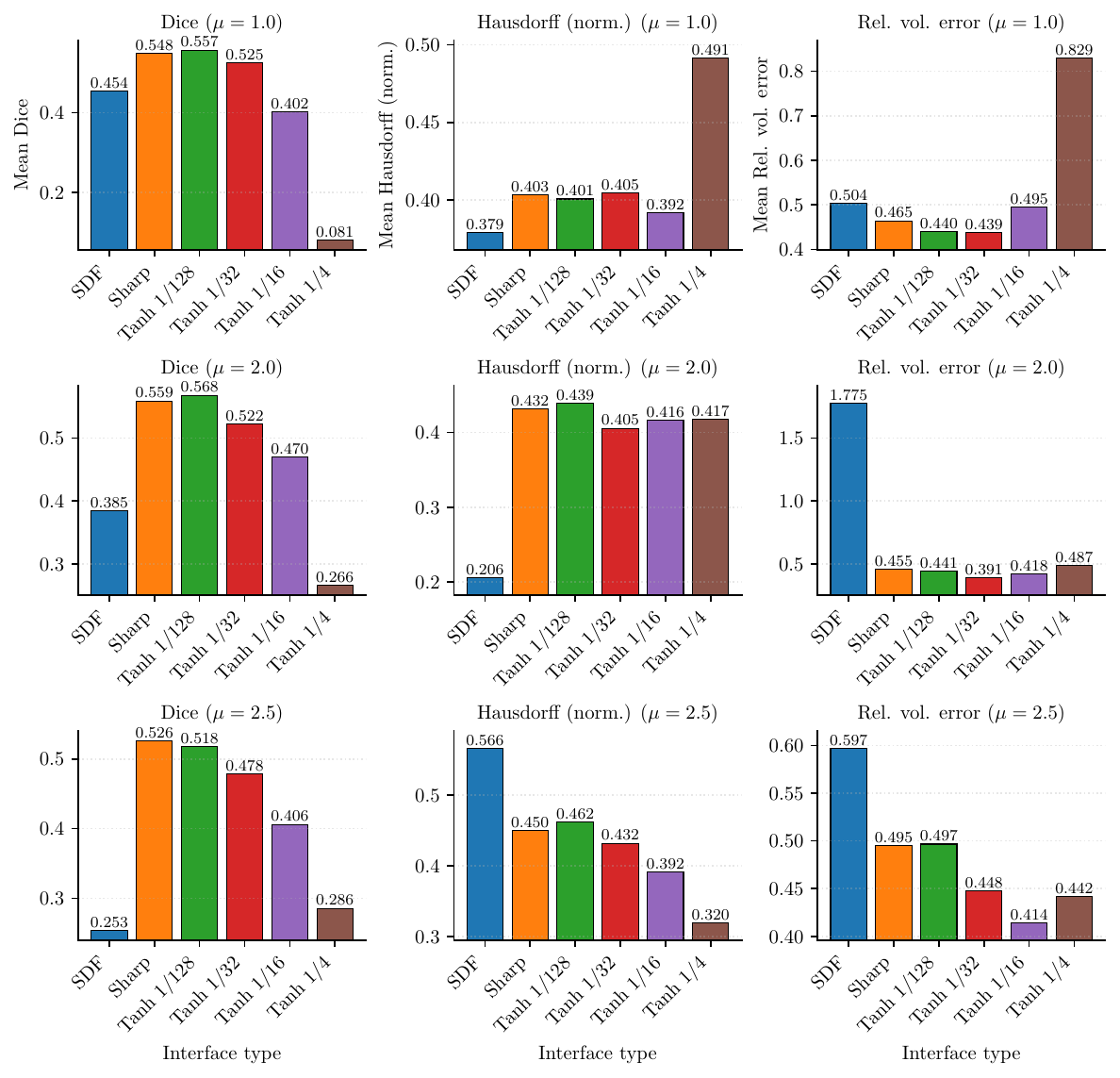}
    \caption{Performance of each interface representation on the synthetic datasets, using a linear autoencoder model. See Figure \ref{fig:comparison_hit_abs} for corresponding results from the nonlinear convolutional autoencoder model.}
    \label{fig:pca}
\end{figure}


\end{document}